\documentclass[showpacs,preprintnumbers,amsmath,amssymb,superscriptaddress]{revtex4}
\usepackage{epsfig}
\usepackage{psfrag}
\usepackage{amsfonts}
\usepackage{graphicx}
\usepackage{dcolumn}
\usepackage{bm}

\begin{document}


\title{Lack of Debye and Meissner screening in strongly magnetized quark matter at intermediate densities }

\author{Bo  Feng}
\affiliation{
School of Physics, Huazhong University of Science and Technology, Wuhan 430074, China
}

\author{Efrain J. Ferrer}
\affiliation{Department of Physics and Astronomy, University of Texas Rio Grande Valley, 1201 West University Dr., Edinburg, TX 78539 and CUNY-Graduate Center, New York 10314, USA}

\author{Israel Portillo}
\affiliation{Department of Physics, University of Houston, Houston, TX 77204, USA}


\date{\today}

\begin{abstract}
We study the static responses of cold quark matter in the intermediate baryonic density region (characterized by a chemical potential $\mu$) in the presence of a strong magnetic field. We consider in particular, the so-called Magnetic Dual Chiral Density Wave (MDCDW) phase,
which is materialized by an inhomogeneous condensate formed by a particle-hole pair. It is shown, that the MDCDW phase is more stable in the weak-coupling regime than the one considered in the magnetic catalysis of chiral symmetry braking phenomenon (MC$\chi$SB) and even than the chiral symmetric phase that was expected to be realized at sufficiently high baryonic chemical potential. The different components of the photon polarization operator of the MDCDW phase in the one-loop approximation are calculated. We found that in the MDCDW phase there is no Debye screening neither Meissner effect in the lowest-Landau-level approximation. The obtained Debye length depends on the amplitude $m$ and modulation $b$ of the inhomogeneous condensate and it is only different from zero if the relation $| \mu -b| > m$ holds. But, we found that in the region of interest this inequality is not satisfied. Thus, no Debye screening takes place under those conditions. On the other hand, since the particle-hole condensate is electrically neutral, the U(1) electromagnetic group is not broken by the ground state and consequently there is no Meissner effect. These results can be of interest for the astrophysics of neutron stars. 
\end{abstract}

\maketitle

\section{Introduction}

It is well known that quantum chromodynamics (QCD) has a rich phase structure. In its usual temperature versus baryon-number density phase map, color superconductivity (CS) is the well established ground state in the asymptotically large density and low temperature region. This phase is characterized by the formation of quark-quark pairs analogous to the BCS pairs of conventional electronic superconductivity. While on the other extreme of low-density/low-temperature region, the quarks are confined into hadrons having large masses produced by the  breaking of chiral symmetry due to the chiral homogeneous condensate formed by quark-antiquark pairs.  In the intermediate density region, however, the energetically favored ground state al low temperature remains murky since neither perturbative QCD nor lattice calculations is applicable in that region. 

Nevertheless, various QCD effective model studies, as well as QCD calculations in the large $N_c$ limit indicate that between the hadronic phase and the more energetic favored superconducting phase (the so called CFL  phase) there would be some intermediate states characterized by inhomogeneous particle-hole condensates, in which the pairs carry a total finite momentum \cite{Large-Nc}-\cite{Landau-Instability}.  Although these studies suggest that the inhomogeneous phases might be unavoidable  in the regions of intermediate temperature and density, the circumstances are still involved due to the fact that the pairing energies between particle-particle, particle-antiparticle and particle-hole are comparable in those regions.   A systematic and complete investigation to determine the most energetically favored state under different conditions is still open and beyond the scope of the present study. Instead, we shall focus our attention on one particular inhomogeneous phase,  the so-called dual chiral density wave (DCDW), whose ground state is characterized by a spatially modulated chiral condensation in both scalar and pseudo-scalar channels.

Notably, in most situations where quark-matter phases can be generated, magnetic fields are usually present. Off-central heavy-ion collisions (HIC) where quark matter degrees of freedom become relevant are known to produce large magnetic fields ($eB \simeq 10^{18}$ G at RHIC, $eB \simeq 10^{19}$ G, at the LHC \cite{Warringa, Skokov}.). Likewise, neutron stars (NS) typically have strong magnetic fields. Estimates based on the scalar virial theorem give inner fields for magnetars of order $10^{18}$ G for nuclear matter \cite{Shapiro} and $10^{20}$ G for quark matter \cite{EOS-Fermions}. Even inner fields, one to three orders of magnitude smaller, would be significant and should not be ignored in NS studies \cite{Hackebill}. The magnetic field can noticeably enhance the window for inhomogeneous phases \cite{Angel, Quiroz, Bo}; and activate attractive channels producing new condensates,  as it occurs with chiral condensate \cite{Quiroz}, color superconductivity \cite{Bo} and quarkyonic matter \cite{Angel}.  

The interplay of magnetic fields with the spatially inhomogeneous chiral condensate is therefore important and worth to be investigated.  Within Nambu-Jona-Lasino (NJL) model, it had been shown that the presence of an external magnetic field favors the formation of spatially inhomogeneous condensate in dense quark matter at low temperature \cite{Klimenko}. Moreover, the constant external magnetic field breaks the rotational symmetry of the system producing Landau momentum quantization, which can give rise to an asymmetric quark energy spectrum in the lowest Landau level (LLL). This asymmetry is basic to originate a nontrivial topology in the system. Recently, it was pointed out that the DCDW phase in a magnetic field is physically distinguishable from the DCDW phase at zero field and thus named as magnetic dual chiral density wave (MDCDW) \cite{E-V-PLB}.  The topology of the MDCDW phase manifests in the effective electromagnetic action by the presence of a dynamical axion field coupled to the electromagnetic field. This coupling in turn leads to several topological effects in dense quark matter, as for instance, an anomalous non-dissipative Hall current, the existence of magnetoelectricity, etc. \cite{E-V-NPB}.

Here the following comment is in order. It is well-known that single-modulated phases in three spatial dimensions are unstable against thermal fluctuations at any finite temperature, a phenomenon known in the literature as Landau-Peierls instability \cite{Landau}. In dense QCD models, the Landau-Peierls instability has been shown to occur in the periodic real kink crystal phase \cite{Hidaka}; in the DCDW phase \cite{Lee}; and in the quarkyonic phase \cite{Pisarski}. The instability signals the lack of long-range correlations at any finite temperature and hence the lack of a true order parameter. Only a quasi-long-range order remains in all these cases, a situation that resembles what happens in smectic liquid crystals \cite{Luban}.
In \cite{Landau-Instability}, the stability of the MDCDW phase against thermal fluctuations in the region of relevance for NS applications was investigated. There, it was shown that a background magnetic field introduces new structures in the system  that are consistent with the symmetry group that remains after the explicit breaking of the rotational and isospin symmetries by the magnetic field. The new terms not only modify the condensate minimum equations, but they also lead to a linear, anisotropic spectrum of the thermal fluctuations, which lacks soft transverse modes. Soft transverse modes are the essence of the Landau-Peierls instability because they produce infrared divergencies in the mean square of the fluctuation field that in turn wipe out the average of the condensate at any low temperature. Since this does not happen in the MDCDW phase, there exists the possibility that the MDCDW long-range order may remain stable within a range of temperatures feasible for this phase to be stable in the NS core. 

Another important magnetic effect in quark matter is that a magnetic field helps the condensation of the homogeneous chiral condensate by increasing the population of the quarks in the LLL, which are then closer to the antiquarks in the Dirac sea, a phenomenon that has been called magnetic catalysis of chiral symmetry breaking (MC$\chi$SB)  \cite{MC}-\cite{Paraelectricity}. The MC$\chi$SB is a universal phenomenon that takes place in any relativistic theory of interactive massless fermions in a magnetic field, and it has been proposed as the mechanism explaining various effects in quasiplanar condensed matter systems \cite{7}.

Now, with increasing chemical potential, the energy separation between quarks and anti-quarks increases up to a point where it is no longer energetically favorable  to excite antiquarks all the way from the Dirac sea to be paired with the quarks at the Fermi surface. When this happens, various possibilities are opened: Either no condensate is favored, and the chiral symmetry is restored; or quarks and holes near the Fermi surface pair with parallel momenta, giving rise to inhomogeneous chiral condensates; or quark may even pair with quarks through an attractive channel at the Fermi surface to form a CS phase that ultimately may be inhomogeneous. Thus, in the presence of a magnetic field, there will be a competition between the strong field effect that forces the quarks to be in the LLL near the antiquarks with whom to form the homogeneous chiral pair and the chemical potential that opens the gap between the quarks on the Fermi surface and the antiquarks in the Dirac see, so favoring the formation of the inhomogeneous condensate of the particle-hole pair sitting on the Fermi surface. 

An interesting question that we want to investigate in this paper is how this competition between the MC$\chi$SB phenomenon and the formation of the MDCDW condensate at weak coupling takes place in the dense region to determine the more energetically favored phase.

Another important feature, which is worthy to study in this context is that, depending on the characteristics of the ground state, quark matter can have different electric screening properties. For example, it is noticeable that finite density, which usually leads to Debye screening in systems of free fermions \cite{Fradkin}, fails to produce the same effect neither in the Color-Flavor-Locked (CFL) phase  \cite{Debye-CFL}, nor in the Magnetic-Color-Flavor-Locked (MCFL)  \cite{Debye-MCFL} phase of CS, because no infrared electric screening can be produced by diquark condensates that are neutral.  Let us recall that even though the original electromagnetic $U(1)_{em}$ symmetry is broken by the formation of quark Cooper pairs in the CFL phase \cite{CFL-Rotated-1}  of CS, a residual $\widetilde{U}(1)$ symmetry still remains. The massless gauge field associated with this symmetry is given by the linear combination of the conventional photon field and the $8$th-gluon field \cite{CFL-Rotated-1, CFL-Rotated-2}, $\widetilde{A}_{\mu}= \cos \theta A_{\mu}-\sin \theta G^8_{\mu}$. The field $\widetilde{A}_\mu$ plays the role of an in-medium or rotated electromagnetic field. Therefore, a magnetic field associated with $\widetilde{A}_\mu$ can penetrate the CS without being subject to the Meissner effect, since the color condensate is neutral with respect to the corresponding rotated electric charge. A similar residual electromagnetic group also remains in the 2SC phase of CS \cite{Bailin}. In this paper, we shall study the screening properties of the MDCDW phase in a strong magnetic field validating the LLL approximation at weak coupling. 

The knowledge of the screening effect in a dense medium is important for astrophysics. Since many year ago, gravitationally bound system under electric fields has been a topic of investigation \cite{E-NS}. It has been found that any bound system whose size is smaller than the Debye length of the surrounding media can be electrified by the escape of electrons, which being the lighter particles in the system are displaced to the star surface. This electron motion is stopped by the electric field created by the corresponding charge separation between the electrons and the other heavier charged particles existing in the interior as protons, up quarks, etc. The effects of the induced electric field may become important to the structure of the star. Thus, to know how this inner field can be screened by the stellar medium is an important ingredient to determine the general setting of this interesting problem. In the same footing, to know if a magnetic field can be screened in the stellar medium is of major importance.

The rest of the paper is organized as follows: in Section II, we introduce the MDCDW model and discuss the validity of the LLL-approximation at strong magnetic-fields in the weak-coupling regime for the density domain of interest.  In Section III, comparing the thermodynamic potentials of the MDCDW and MC$\chi$SB  phases, we prove that at weak-coupling, the MDCDW phase is energetically favored in the whole density domain. In Section IV, we calculate the photon polarization operator in the one-loop approximation and in the strong-field limit for the MDCDW phase. From its infrared limit we obtain the Debye and Meissner masses. The Debye mass has a dynamical nature depending on the condensate parameters. Then, we show that in the density region of interest the Debye length is zero. Hence, there is no electrostatic screening in this medium under the presence of the particle-hole pairs. We also show that there is no Meissner screening in this medium. In Section V, we summarize the outcomes of this paper and give the concluding remarks. Finally,  some calculation details  are presented in the Appendices.

\section{Strong magnetic field limit in the weak-coupled MDCDW quark matter phase}

We shall consider in this paper a two flavor NJL model of massless quarks in the presence of an external constant and uniform magnetic field. The Lagrangian reads
\begin{equation}
{\cal L}={\bar\psi}\left[i\gamma^\mu(\partial_\mu+iQA_\mu)+\gamma^0\mu\right]\psi+G\left[\left({\bar\psi \psi}\right)^2+\left({\bar\psi}i\tau^a\gamma^5\psi\right)^2\right],
\end{equation}
with $Q={\rm diag}(e_u,e_d)={\rm diag}(2e/3,-e/3)$ being the charge operator of the quark doublet $\psi^T=(u,d)$, $\mu$ the baryonic chemical potential, and $G$ the coupling constant of the four-fermion interaction. $\tau^a$ with $a=1,2,3$ are the three Pauli matrices. For the external magnetic field we impose the Landau gauge, in which the electromagnetic potential $A^\mu=(0,0,Bx,0)$  corresponds to a constant and uniform magnetic field $B$  in the positive $z$ direction.

The magnetic dual chiral density wave (MDCDW) phase is characterized by the following scalar and pseudoscalar condensates
\begin{equation}
\langle{\bar\psi}\psi\rangle=\Delta\cos q_\mu x^\mu, \ \ \ \ \langle{\bar\psi}i\tau^3\gamma^5\psi\rangle=\Delta\sin q_\mu x^\mu,
\end{equation}
with $\Delta$ the magnitude of the condensate and $q^\mu=(0,0,0,q)$ its modulation, which is taking along the field direction since this configuration minimizes the system energy \cite{Klimenko}. The MDCDW phase has a symmetry different from the DCDW one. The flavor symmetry $SU(2)_L\times SU(2)_R$ of the DCDW phase is reduced to the subgroup $U(1)_L\times U(1)_R$ due to the coupling of quarks to the magnetic field, and more importantly, the presence of the magnetic field induces a non-trivial topology in the MDCDW phase that gives rise to axion electrodynamics \cite{E-V-PLB, E-V-NPB}. 

For the sake of clarity and understanding, we will summarize as follows some of the results that were obtained in other works (see for example Refs. \cite{Klimenko, E-V-PLB, E-V-NPB}).
In the mean field approximation, the Lagrangian of the MDCDW is
\begin{equation}
{\cal L}_{\rm MF}={\bar\psi}\left[i\gamma^\mu(\partial_\mu+iQA_\mu)+\gamma^0\mu-m\left(\cos qz+i\tau^3\gamma^5\sin qz\right)\right]\psi-\frac{m^2}{4G},
\end{equation}
with $m=-2G\Delta$. Now, we use the gauge chiral transformation,
\begin{equation}
\psi\rightarrow  e^{-i\tau^3\gamma^5bz}\psi, \ \ \ \ {\bar\psi}\rightarrow {\bar\psi}e^{-i\tau^3\gamma^5bz},\label{Chiral_Trans}
\end{equation}
with $b=q/2$, to remove the spatial modulation in the mean field Lagrangian and obtain
\begin{equation}
{\cal L}_{\rm MF}={\bar\psi}\left[i\gamma^\mu(\partial_\mu+iQA_\mu)+\gamma^0\mu+\tau^3\gamma^3\gamma^5b-m\right]\psi-\frac{m^2}{4G}.\label{meanfieldlagrangian}
\end{equation}

As it was found in \cite{E-V-PLB, E-V-NPB}, there is a lack of invariance of the path-integral fermion measure under the chiral gauge transformation (\ref{Chiral_Trans}). Thus, in  \cite{E-V-PLB, E-V-NPB} it was used the Fujikawa's method \cite{Fujikawa} together with a representation of the Jacobian in terms of a complete and orthogonal set of eigenfunctions that ensured unitarity via the diagonalization of the fermion action. Considering the appropriate set of eigenfunctions, 
it was found, in a gauge-invariant way, the contribution to the effective action that came from the regularized path integral measure. This contribution turned out to be a chiral anomaly term $(\kappa/4)\theta F^\ast_{\mu\nu} F^{\mu\nu}$, that coupled the axion field $\theta=qz/2$ to the electromagnetic strength tensor $F_{\mu\nu}$ and its dual with coupling $\kappa=\alpha/2\pi$.  Nevertheless, we should notice that for the investigation of the screening effects of this phase, the axion term is not relevant, since it will not contribute to the quark propagator forming the internal lines in the photon polarization operator (see Eq. (\ref{photonselfenergy})), which is the QFT diagram from where the screening effects are investigated..

The effective potential of the system is
\begin{align}
\Omega &=-\frac{1}{\tilde V}\ln {\cal Z},
\end{align}
with $\tilde V$ the four-dimensional volume and
\begin{equation}
{\cal Z}=\int {\cal D}{\bar\psi}(x){\cal D}\psi(x)e^{i\int d^4x\left\{{\bar\psi}\left[i\gamma^\mu(\partial_\mu+iQA_\mu)+\gamma^0\mu+\tau^3\gamma^3\gamma^5b-m\right]\psi-\frac{m^2}{4G}\right\}}.
\end{equation}
Upon a Wick rotation from Minkowski to Euclidean spacetime, the functional $\cal Z$  is just the partition function of the system and $\Omega$ becomes the thermodynamic potential.  Integrating out the fermion field, one obtains
\begin{align}
\Omega={\rm Tr}\ln \left[i\gamma^\mu(\partial_\mu+iQA_\mu)+\gamma^0\mu+\tau^3\gamma^3\gamma^5b-m\right]+\frac{m^2}{4G}.
\end{align}
The trace,  extended to flavor, color and spinor indices,  can be evaluated in the presence of a magnetic field through a standard procedure, as for instance the Ritus' approach \cite{Ritus}. Hence, we obtain
\begin{equation}
\Omega=-\frac{N_c}{2}\int\!\!\!\!\!\!\!\!\sum\frac{d^4p}{(2\pi)^4}{\rm tr}\ln\left[-(p^0)^2+(E_l-\mu)^2\right]+\frac{m^2}{4G},\label{effectivepotential}
\end{equation}
where $N_c$ is the number of color degrees of freedom and the trace $\rm tr$ is extended to flavor indices only.  In (\ref{effectivepotential}), we used the notation $\int\!\!\!\!\!\!\sum{d^4p}= \frac{1}{\beta}\sum_{n=-\infty}^{\infty}\sum_{l=0}^\infty\int dp^2dp^3\sum_{\epsilon=\pm,\varepsilon=\pm}$, with $l$ the Landau-level number and the statistical formulation is introduced following Matsubara's prescription for fermions
\begin{equation}\label{Matsubara}
p^0\rightarrow i\omega_n=\frac{i}{\beta}(2n+1)\pi, \quad \int_{-\infty}^\infty\frac{dp^0}{2\pi} \rightarrow \frac{i}{\beta}\sum_{n=-\infty}^{\infty},
\end{equation}
with $\beta=1/T$. 

The energy spectrum \cite{Klimenko} 
\begin{equation}
E_l=\epsilon\sqrt{\left(\varepsilon\sqrt{m^2+p_3^2}+b\right)^2+2|e_fB|l}, \quad l=1,2,3,...,
\end{equation}
 and
\begin{equation}
E_0=\epsilon\sqrt{m^2+p_3^2}+b, \quad l=0,
\end{equation}
with $\epsilon=\pm, \varepsilon=\pm$, are the eigenvalues of the single particle Hamiltonian that can be read from the Lagrangian (\ref{meanfieldlagrangian}) for each flavor as
\begin{equation}
{\cal H}_f=-i\gamma^0\gamma^i\left [\partial_i+ie_fA_i+ib\gamma^3\gamma^5 {\rm sgn}(e_f)\right ].
\end{equation}
Here $e_f$ with $f=u,d$ are the electric charges of the $u$ and $d$ quarks respectively.
 
The thermodynamic potential is then given by
\begin{equation}
\Omega=-\frac{N_c}{2}\sum_{f=u,d}\frac{|e_fB|}{\beta}\sum_n\sum_{l,\varepsilon,\epsilon}\int\frac{dp_3}{(2\pi)^2}\ln\left[-(i\omega_n)^2+(E_l-\mu)^2\right]+\frac{m^2}{4G}.
\end{equation}
The sum over Matsubara frequencies $\omega_n$ can be readily carried out and one obtains
\begin{equation}
\Omega=-\frac{N_c}{2}\sum_{f=u,d}|e_fB|\sum_{l,\varepsilon,\epsilon}\int\frac{dp^3}{(2\pi)^2}\left[|E_l-\mu|+\frac{2}{\beta}\ln\left(1+e^{-\beta|E_l-\mu|}\right)\right]+\frac{m^2}{4G}.
\end{equation}

\begin{figure}
	\includegraphics[height=13cm]{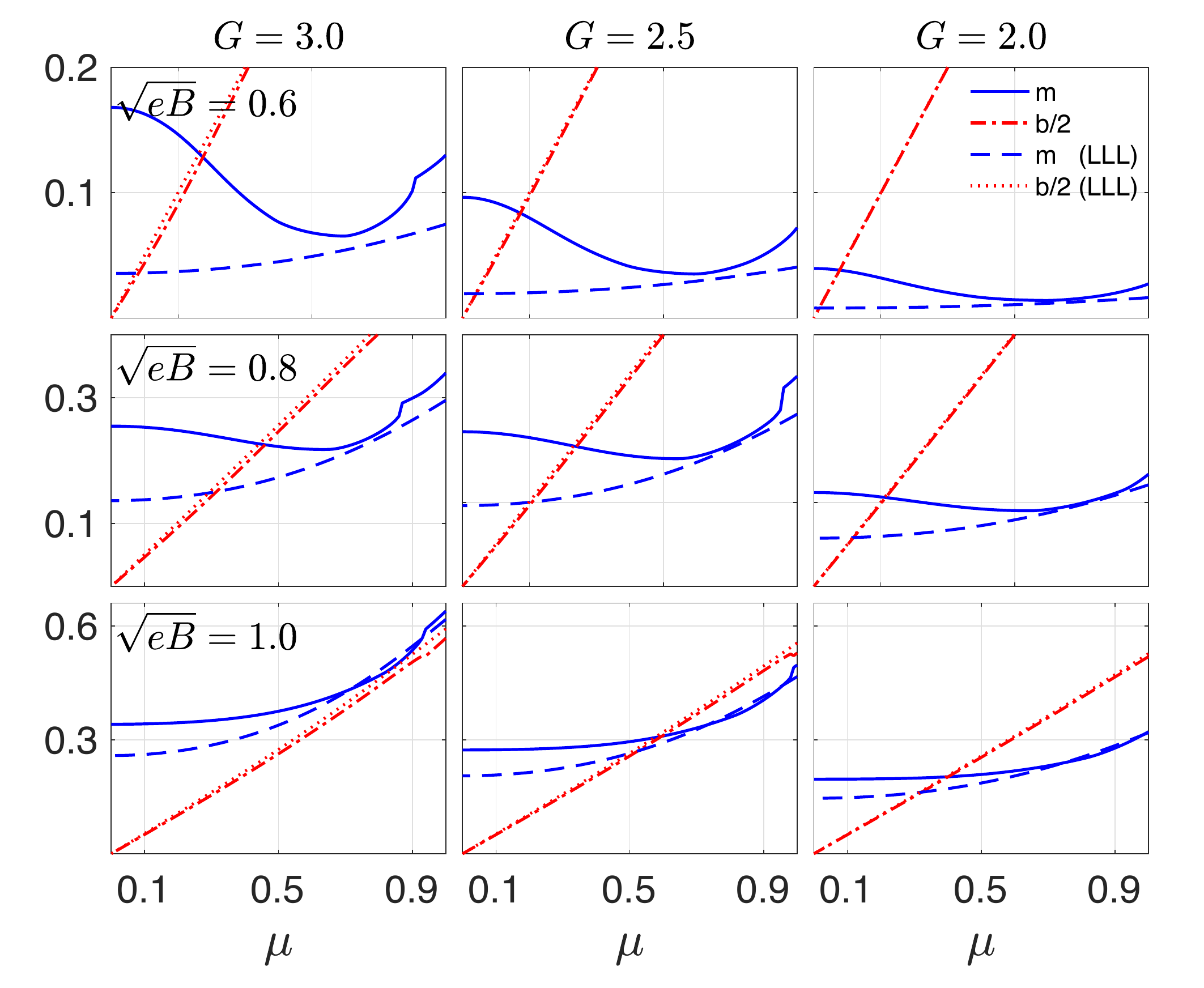}
	\caption{ (Color online) The solutions to the gap equations of the MDCDW phase versus baryonic chemical potential at different values of the coupling constant ($G= 2, 2.5, 3$) and at magnetic-field values ($\sqrt{eB}=0.6, 0.8, 1.0$). The solutions for $m$ and $b/2$ are obtained summing in all Landau levels and for $m$(LLL) and $b/2$(LLL) by only taking into account the lowest Landau level.} \label{Fig-1}
\end{figure}

Obviously, the vacuum part in the thermodynamic potential is UV divergent and needs to be regularized. Using the proper-time regularization, the regularized potential reads \cite{Klimenko}
\begin{equation} \label{thermo-pot}
\Omega=\Omega_{\rm vac}+\Omega_{\rm anom}+\Omega_{\mu}+\Omega_T+\frac{m^2}{4G},
\end{equation}
with
\begin{equation}
\Omega_{\rm vac}=N_c\frac{1}{4\sqrt{\pi}}\sum_{f=u,d}|e_fB|\sum_{l,\varepsilon,\epsilon}\int\frac{dp^3}{(2\pi)^2}\int_{1/\Lambda^2}^\infty\frac{ds}{s^{3/2}}e^{-sE_l^2},
\end{equation}
\begin{equation}\label{Omega_anom}
\Omega_{\rm anom}=-N_c\frac{b\mu}{2\pi^2}\sum_{f=u,d}|e_fB|,
\end{equation} 
\begin{equation} \label{omegamu}
\Omega_\mu=-\frac{1}{2}\sum_{f=u,d}|e_fB|\sum_{l,\varepsilon,\epsilon}\int\frac{dp^3}{(2\pi)^2}\left(|E_l-\mu|-|E_l|\right),
\end{equation}
and
\begin{equation}
\Omega_T=-\sum_{f=u,d}|e_fB|\sum_{l,\varepsilon,\epsilon}\int\frac{dp^3}{(2\pi)^2}\ln\left(1+e^{-\beta|E_l-\mu|}\right).
\end{equation}

The anomalous contribution $\Omega_{\rm anom}$ comes from the regularization of the LLL part ensuring that the thermodynamic potential is independent of $b$ when $m=0$ \cite{Klimenko}.

\begin{figure}
	\includegraphics[height=10cm]{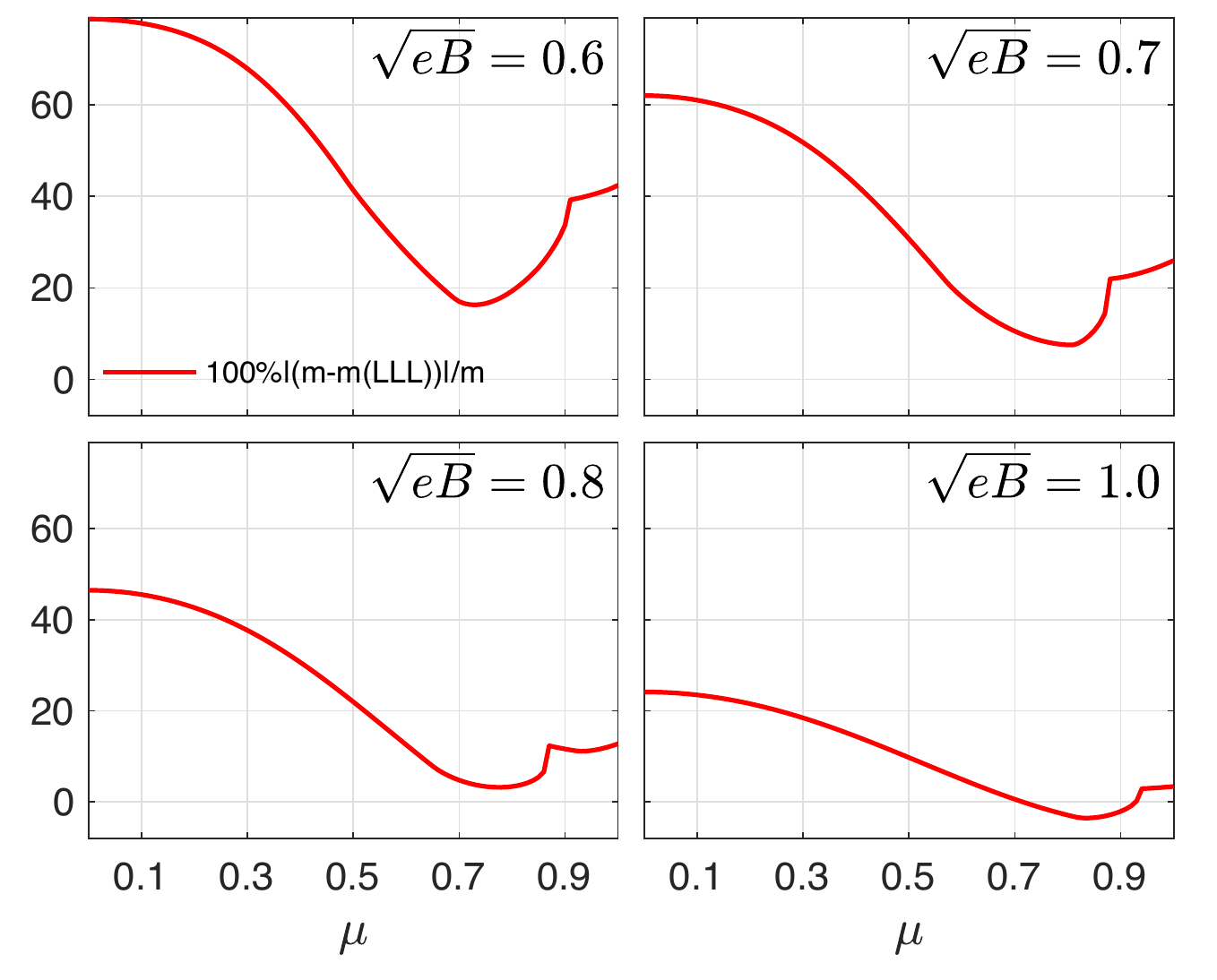}
	\caption{Percentage deviation, $[(m-m(LLL))/m]\times 100 \%$, versus chemical potential for four magnetic-field values and at a coupling constant G=3. The deviation is taking between the induced mass calculated summing in all LL's, $m$, and only considering the LLL, m(LLL). }\label{Fig-Percentage-Validity}
\end{figure}

The order parameters $m$ and $b$ should be determined from the gap equations 
\begin{equation}\label{Gap-Eq}
\frac{\partial\Omega}{\partial m}=\frac{\partial\Omega}{\partial b}=0,
\end{equation}
what indicates their dynamical origin.

In what follows, we rescale all dimensional quantities with the cutoff $\Lambda$, as for instance, $m/\Lambda, b/\Lambda$, $G\Lambda^2$, etc. In order to keep a simplified notation, we simply use the same symbols as the original ones to represent the new dimensionless quantities. 

We are interested in comparing the properties of the MC$\chi$SB phenomenon with those of the MDCDW phase. As it is known, in the MC$\chi$SB it was considered a weak coupling massless fermion theory in the presence of a magnetic field that is the leading parameter. In those circumstances, the fermions were mostly confined to the LLL, and the study of the chiral condensation phenomenon was limited to solve the gap equations for a thermodynamic potential in the LLL-approximation. Thus, a first step in our investigation will be to determine the range of parameters where the same approximation will be valid in the MDCDW phase. 

\begin{figure}
	\includegraphics[height=10cm]{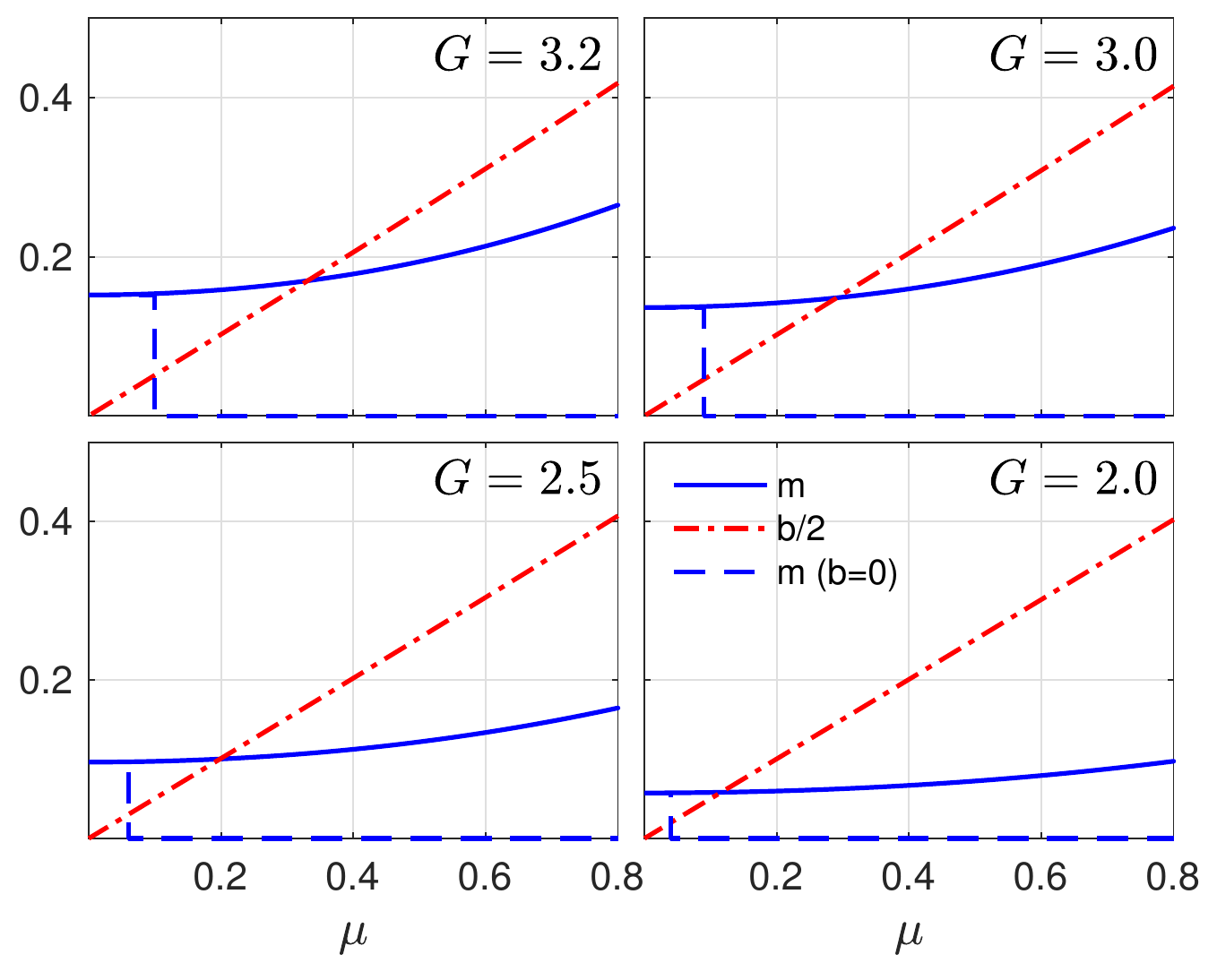}
	\caption{(Color online) The solutions to the gap equations of the MDCDW ($m$ and $b$) and MC$\chi$SB ($m(b=0)$) phases versus baryonic chemical potential at different values of the coupling constant ($G= 2, 2.5, 3, 3.2$) and at a fixed magnetic field ($\sqrt{eB}=0.8$). The solutions are obtained in the LLL-approximation. }\label{Fig-2}
\end{figure}

In Fig.\ref{Fig-1}, we show the solutions for $m$ and $b$ versus the baryonic chemical potential for different coupling strengths and magnetic-field values. The calculations are implemented for two cases:  for a thermodynamic potential in the LLL approximation and for a thermodynamic potential obtained by summing in all LL's (see Fig. caption). There, we notice that while the modulation $b$ has similar values independent of the used approximation, the condensate amplitude, $m$, coincides only at weak coupling in the region of higher chemical potentials. Also notice that the modulation ($b\neq 0$)  is present in all the $\mu \neq 0$ domain. This is a consequence of the presence of the anomalous contribution (\ref{Omega_anom}).

To determine the validity region of the LLL approximation, in Fig. \ref{Fig-Percentage-Validity} we plotted the percentage deviation, defined as $[(m-m(LLL))/m]\times 100 \%$, to have a better criterium of what is the density region where the approximation is sound. Considering that in a strict way, the magnetic field becomes the scale, i.e. $eB \sim \Lambda^2$, when the LLL approximation is valid, the values in Fig. 2 corresponding to $\sqrt{eB}=1.0$ are the more trustworthy ones. For this magnetic field, we can see from Fig. \ref{Fig-Percentage-Validity} that for a percentage deviation of 20$\%$ the density region is $0.2 \lesssim \mu \lesssim 1$ and for a 10$\%$ is $0.5 \lesssim \mu \lesssim 1$. Decreasing the magnetic field, the density validity regions shrink, moving  to higher values of the chemical potential. In Fig. \ref{Fig-Percentage-Validity} we considered only one coupling constant value ($G=3$), but we checked that changing $G$ among the value set considered in Fig. \ref{Fig-1} the valid density intervals remain practically the same.

Then, contrary to what occurred in the  MC$\chi$SB case, when working in the MDCDW phase and using the LLL approximation, we should be carful in specifying the valid density region where the results are reliable. The situation here is different to what happens for example in QED under a strong magnetic field at zero density, where in the strong-field limit it is equivalent to sum in all Landau levels or to work in the LLL \cite{Angel-2}.

\begin{figure}
	\includegraphics[height=10cm]{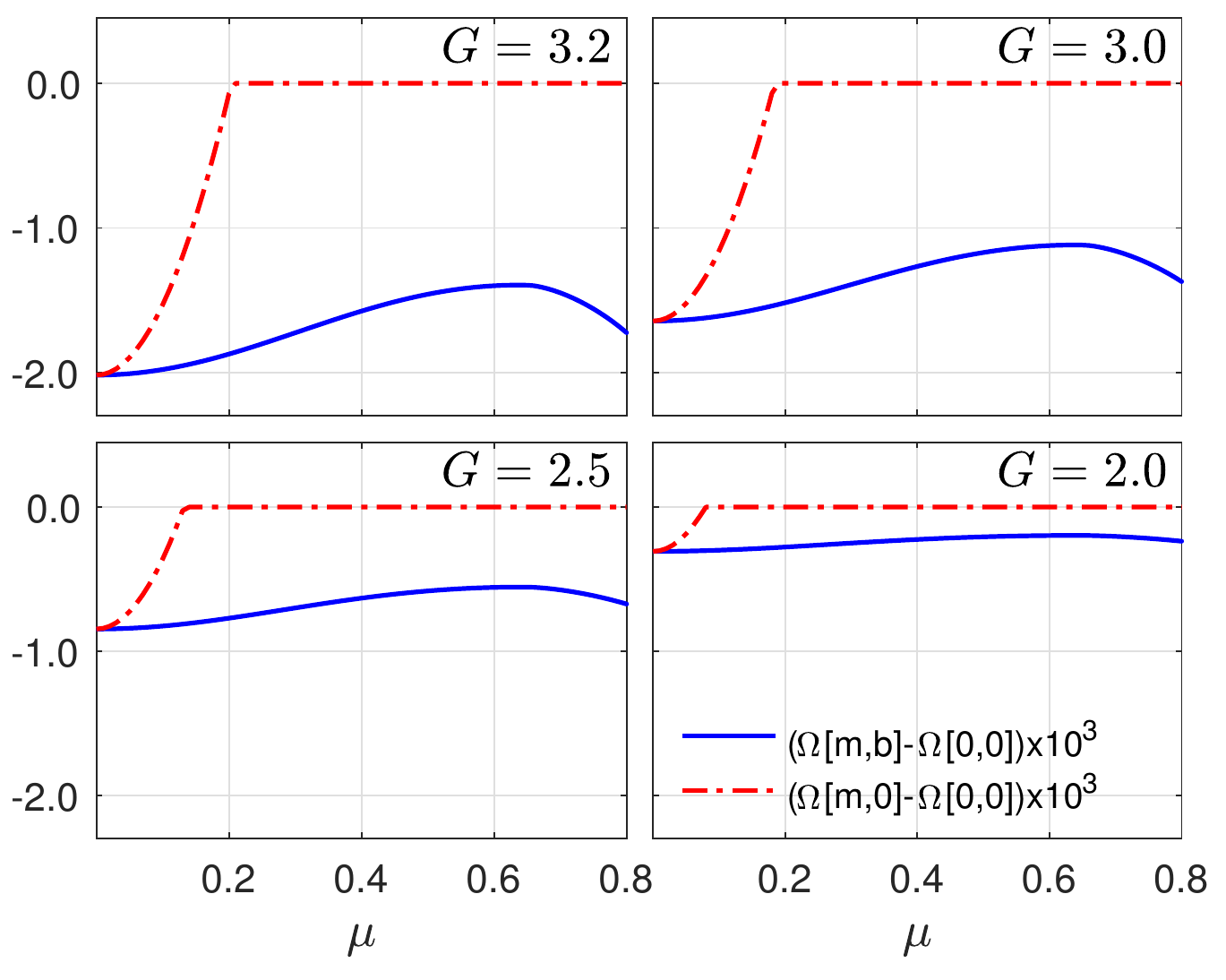}
	\caption{(Color online) Thermodynamic potential evaluated in the gap solutions of the MDCDW phase, $\Omega(m, b)$, minus its zero-condensate value, $\Omega(m=0, b=0)$ (solid line), and for the MC$\chi$SB phase, $\Omega(m, b=0)-\Omega(m=0, b=0)$ (interrupted line), versus the baryonic chemical potential at different values of the coupling constant ($G= 2, 2.5, 3, 3.2$) for $\sqrt{eB}=0.8$ and summing in all Landau levels.}\label{Fig-3}
\end{figure}

Finally, the following technical comment is in order. It should be noticed that the sum in Landau levels in the many-particle  thermodynamic potential at zero temperature (\ref{omegamu}) does not require the introduction of a cutoff parameter $\Lambda$, as it was the case in color superconductivity, where a soft-cutoff was necessary (see for example Eq. (6) in Ref. \cite{Paulucci}). In the present case, expression (\ref{omegamu}) is automatically regularized by the baryonic chemical potential, which enforces the Fermi momentum as a natural cutoff at zero temperature, as it is explicitly seen in Eq. (73) of Ref. \cite{E-V-NPB}.

\section{Magnetic catalysis of chiral symmetry breaking versus MDCDW phase of dense quark matter in a strong magnetic field}

As we already pointed out, in the MC$\chi$SB phenomenon at finite density, we have two competing elements: the strong magnetic field and the particle density. While the magnetic field acts in favors of the condensation by increasing the population of the quarks in the LLL that are the ones closed in the energy-momentum space to the antiquarks with which to form the chiral condensate; the density tends to break the chiral pair by increasing the energy gap between the Fermi surface and the Dirac  sea. A valid question is then, what will be the effect of these two ingredients in the particle-hole condensate of the MDCDW phase? The first noticeable difference is that since the particle and the hole forming the pair are both sitting on the Fermi surface, the chemical potential, whose function is to increase the Fermi sphere radius, will not tend to break the pair since it will not produce an energy gap between the two pair components, as it was done in the MC$\chi$SB case between the particle and the antiparticle.

In Fig.\ref{Fig-2}  it is plotted the MC$\chi$SB and MDCDW condensate parameters versus the baryonic chemical potential  for a strong magnetic field value at different coupling constants, all in the range of weak couplings (the critical value for the weak regime in the MDCDW phase is $G_c=3.27$ \cite{Klimenko}). We notice that in each case there is a critical value of the chemical potential, $\mu_c$, where the chiral condensate of the MC$\chi$SB phase vanishes. By increasing the strength of the coupling constant, $\mu_c$ increases, but always having values that satisfy $\mu_c \lesssim m(\mu=0)$, where  $m(\mu=0)$ is the particle mass at zero chemical potential. On the contrary, the amplitude and modulation of the particle-hole condensates in the region of intermediate densities where the LLL approximation is valid increase with the chemical potential. These results validate the physical arguments we exposed before. 

Once we know how the condensates of the two phases behave in the region of interest, we can answer another important question: what of the two weak-coupled phases is the more energetically favored at strong magnetic fields in the intermediate density region? To answer this question, we calculate the corresponding thermodynamic potential for each phase evaluated in the minimum solutions of the condensates obtained for each value of chemical potential at the strong magnetic field value of $\sqrt{eB}=0.8$ and at different values of the coupling constant $G$. Note that in this calculations, whose results are given in Fig. \ref{Fig-3}, the sum in all LL's was carried out.
We can see from the plots in Fig. \ref{Fig-3} that in all cases, the MDCDW phase, having lower values of its thermodynamic potential, is more energetically favored than the MDCDW phase. Notice that in Fig. \ref{Fig-3}, in order to make a sharper distinction between the two cases, we subtracted from each thermodynamic potential the non-condensate dependent part $\Omega(0,0)$. 

Moreover, we can observe from Fig. \ref{Fig-3} that strengthening the interaction between the pairs' ingredients by increasing the coupling constant $G$, the difference between the two thermodynamic potentials is enlarged, which implicates that the MDCDW phase becomes more stable under the given conditions than the MC$\chi$SB phase. 

Finally, we see from Fig. \ref{Fig-3} that in the region where the potential $\Omega(m,0)-\Omega(0,0)$ of the MC$\chi$SB is different from zero (i.e. when $m\neq 0$), its value is always larger than that of the MDCDW phase. This can be connected to the presence of the anomalous contribution (\ref{Omega_anom}) in the thermodynamic potential of the MDCDW phase. Thus, we conclude that the particle-hole inhomogeneous condensate prevails at any given density value in the weak-coupling regime under a strong magnetic field. Therefore, the MC$\chi$SB phenomenon, as it is known to be characterized by an homogeneous chiral condensate, is only energetically realizable at $\mu = 0$.

\section{Photon polarization operator in the strongly magnetized MDCDW phase, Debye and Meissner masses}

\subsection{Photon polarization operator}
In this section, we shall calculate the one-loop photon polarization operator in the MDCDW phase in a strong magnetic field using the LLL-approximation in the quark propagators.   The photon polarization operator is a second-rank tensor that in the above approximation reads
\begin{equation}
\Pi^{\mu\nu}(p^\parallel)=\sum_f\Pi_f^{\mu\nu}(p^\parallel)=\sum_f N_ce_f^2|e_fB|\int\frac{d^2 q^\parallel}{(2\pi)^3}{\rm Tr}\left[\Delta({\rm sgn}(e_f))\gamma^{\mu\parallel} S^{LLL}_f({\tilde q}^\parallel-{\tilde p}^\parallel)\Delta({\rm sgn}(e_f))\gamma^{\nu\parallel} S^{LLL}_f({\tilde{q}}^\parallel)\right],\label{photonselfenergy}
\end{equation}
where $\Delta(\pm)=(1\pm i\gamma^1\gamma^2)/2$ are the spin projectors and $S^{LLL}_f$ is the quark propagator of the MDCDW phase in the LLL-approximation for one flavor \cite{E-V-NPB},
\begin{equation}
S^{LLL}_f({\tilde q}^\parallel)=\frac{{\tilde q}^{\mu\parallel}_+\gamma_\mu^\parallel +m}{({\tilde q}_+^\parallel)^2-m^2}\Delta(+)+\frac{{\tilde q}^{\mu\parallel}_-\gamma_\mu^\parallel +m}{({\tilde q}_-^\parallel)^2-m^2}\Delta(-), \label{PO1}
\end{equation}
with ${\tilde q}^{\mu\parallel}_{\pm}=(q^0-\mu\pm {\rm sgn}(e_f)b,0,0,q^3)$ and $\gamma^{\mu\parallel}=(\gamma^0,0,0,\gamma^3)$. 

\subsection{Debye mass}

As known, the Debye mass is obtained from the $00-$component of the photon polarization operator. Thus, specializing the Lorentz indices in (\ref{photonselfenergy}) as $\mu=\nu=0$ and using the Matsubara technique with the prescription given in (\ref{Matsubara}), we obtain after taken the trace
\begin{align}
 \Pi_f^{44}(p^\parallel)&=2N_ce_f^2 \frac{|e_fB|}{\beta}\sum_{q_4}\int\frac{dq^3}{(2\pi)^2}\frac{(iq^4-\tilde\mu)(iq^4-\tilde\mu-ip^4)+q^3(q^3-p^3)+m^2}{\left[(iq^4-\tilde\mu-ip^4)^2-\epsilon_{q^3-p^3}^2\right][(iq^4-\tilde\mu)^2-\epsilon_{q^3}^2]},\label{00component}
\end{align}
with $\epsilon_{q}=\sqrt{q^2+m^2}$ and $\tilde\mu=\mu-b$. The Matsubara sum can be carried out by using the contour integral and, as usual, the results can be divided into the QFT-vacuum part, $\Pi^{00}_{\rm vac}$, and the statistical part, $ \Pi^{00}_{\rm med}$, which after making the analytic continuation from Euclidean to Minkowski space  reads
\begin{equation}
 \Pi_f^{00}= \Pi^{00}_{\rm vac}+ \Pi^{00}_{\rm med},
\end{equation}\label{Pi}
with
\begin{align}
\Pi^{00}_{\rm vac}(p^\parallel)
=-N_ce_f^2 |e_fB|\int\frac{dq^3}{(2\pi)^2}\left(\frac{1}{\epsilon_{q^3}}+\frac{1}{\epsilon_{q^3-p^3}}\right)\frac{\epsilon_{q^3}\epsilon_{q^3-p^3}-q^3(q^3-p^3)-m^2}{(p^0)^2-(\epsilon_{q^3}+\epsilon_{q^3-p^3})^2},\label{vacpart}
\end{align}
and
\begin{align}
	\nonumber \Pi_{\rm med}^{00}(p^\parallel)= 2N_ce^2 |eB|\int\frac{dq^3}{(2\pi)^2}&\left[\frac{(p^0+\epsilon_{q^3-p^3})\epsilon_{q^3-p^3}+q^3(q^3-p^3)+m^2}{2\epsilon_{q^3-p^3}(p^0+\epsilon_{q^3-p^3}+\epsilon_{q^3})(p^0+\epsilon_{q^3-p^3}-\epsilon_{q^3})} n_F(\epsilon_{q^3-p^3}+\tilde\mu)\right.\\
	\nonumber &+\frac{-(p^0-\epsilon_{q^3-p^3})\epsilon_{q^3-p^3}+q^3(q^3-p^3)+m^2}{2\epsilon_{q^3-p^3}(p^0-\epsilon_{q^3-p^3}+\epsilon_{q^3})(p^0-\epsilon_{q^3-p^3}-\epsilon_{q^3})} n_F(\epsilon_{p^3-q^3}-\tilde\mu)\\
	\nonumber &+\frac{-(p^0-\epsilon_{q^3})\epsilon_{q^3}+q^3(q^3-p^3)+m^2}{2\epsilon_{q^3}(p^0-\epsilon_{q^3-p^3}-\epsilon_{q^3})(p^0+\epsilon_{q^3-p^3}-\epsilon_{q^3})} n_F(\epsilon_{q^3}+\tilde\mu)\\
 &\left.+\frac{-(p^0+\epsilon_{q^3})\epsilon_{q^3}+q^3(q^3-p^3)+m^2}{2\epsilon_{q^3}(p^0-\epsilon_{q^3-p^3}+\epsilon_{q^3})(p^0+\epsilon_{q^3-p^3}+\epsilon_{q^3})} n_F(\epsilon_{q^3}-\tilde\mu)\right].\label{medpart1}
\end{align}
Here, $n_F(z)=1/(1+e^{\beta z})$ is the Fermi-Dirac distribution function. In general, as seen from (\ref{medpart1}), the statistical part depends on temperature and chemical potential. In order to carry out then the infrared limit ($p_0=0, p^3\rightarrow 0$), needed for static responses, it is convenient to rewrite (\ref{medpart1}) in the following way
\begin{align}
\nonumber \Pi^{00}_{\rm med}(p^\parallel)=& 2N_ce_f^2|e_fB|\int\frac{dq^3}{(2\pi)^2}\\
\nonumber &\times\left\{\frac{1}{4}\left(1+\frac{q^3(q^3-p^3)+m^2}{\epsilon_{q^3}\epsilon_{q^3-p^3}}\right)\left[\frac{n_F(\epsilon_{q^3-p^3}+\tilde\mu)-n_F(\epsilon_{q^3}+\tilde\mu)}{p^0-\epsilon_{q^3}+\epsilon_{q^3-p^3}}-\frac{n_F(\epsilon_{q^3-p^3}-\tilde\mu)-n_F(\epsilon_{q^3}-\tilde\mu)}{p^0+\epsilon_{q^3}-\epsilon_{q^3-p^3}} \right]\right.\\
&+\left.\frac{1}{4}\left(1-\frac{q^3(q^3-p^3)+m^2}{\epsilon_{q^3}\epsilon_{q^3-p^3}}\right)\left[\frac{n_F(\epsilon_{q^3-p^3}+\tilde\mu)+n_F(\epsilon_{q^3}-\tilde\mu)}{p^0+\epsilon_{q^3}+\epsilon_{q^3-p^3}}-\frac{n_F(\epsilon_{q^3-p^3}-\tilde\mu)+n_F(\epsilon_{q^3}+\tilde\mu)}{p^0-\epsilon_{q^3}-\epsilon_{q^3-p^3}} \right]\right\}.\label{medpart2}
\end{align}

In the present situation, we are interested only in the screening effect in a cold-dense medium, which are the conditions that better simulates what happens in a NS's core where $\mu\gg T$. Hence, we will take the zero-temperature limit of $\Pi^{00}_{\rm med}$. Moreover, in order to investigate the electrostatic screening in that medium,  we need to search for the infrared behavior of $ \Pi_f^{00}(p^\parallel)$. In statistics, it is well known that the limits $p_0\rightarrow 0$ and ${\bf p}\rightarrow 0$ do not commute. But, the prescription to get the medium's static response is to take first $p_0=0$ and then ${\bf p}\rightarrow 0$ \cite{Linde}.

For the vacuum part, we can regularize it by Pauli-Villars (PV) scheme, which removes all UV divergences and maintains gauge invariance,
\begin{align}
	\Pi^{00}_{\rm vac}(p^\parallel)
	=-N_ce_f^2 |e_fB|\sum_{s=0}^SC_s\int\frac{dq^3}{(2\pi)^2}\left(\frac{1}{\epsilon_{q^3,M_s}}+\frac{1}{\epsilon_{q^3-p^3,M_s}}\right)\frac{\epsilon_{q^3,M_s}\epsilon_{q^3-p^3,M_s}-q^3(q^3-p^3)-M_s^2}{(p^0)^2-(\epsilon_{q^3,M_s}+\epsilon_{q^3-p^3,M_s})^2},
\end{align}
with $\epsilon_{q^3,M_s}=\sqrt{(q^3)^2+M_s^2}$, $M_0=m$, $C_0=1$, and the auxiliary masses $M_{s\ne 0}\rightarrow \infty$ after the integration. The coefficients $C_s$ are judiciously chosen so as to remove some of the singularities in the vacuum part. The strongest singularity was eliminated by imposing $\sum_sC_s=0$ and other singularities, if exist, would disappear if the condition $\sum_sC_sM_s^2=0$ holds \cite{PV}.   In the infrared limit we have 
\begin{align}
	\Pi^{00}_{\rm vac}(p^0=0, p^3\rightarrow 0)
	=N_ce_f^2 |e_fB|\sum_{s=0}^SC_s\int\frac{dq^3}{(2\pi)^2}\left[-\frac{q^3p^3}{\epsilon_{q^3,M_s}^3}+{\cal O}((p^3)^2)\right].
\end{align}
As expected, the vacuum-part contribution vanishes in the leading order. 

 For the medium part, we start from the zero-temperature limit expression with $p^0=0$, 
\begin{align}
\nonumber \Pi^{00}_{\rm med}(p^0=0, p^3)=&\frac{N_ce_f^2 |e_fB|}{2}
\int\frac{dq^3}{(2\pi)^2}\left[\left(1+\frac{q^3(q^3-p^3)+m^2}{\epsilon_{q^3}\epsilon_{q^3-p^3}}\right)\frac{\theta(\tilde\mu-\epsilon_{q^3-p^3})-\theta(\tilde\mu-\epsilon_{q^3})}{\epsilon_{q^3-p^3}-\epsilon_{q^3}} \right.\\
&\ \ \ \ \ \ \ \ \ \ \ \ \ \ \ \ \ \ \ \ \ \ \  +\left.\left(1-\frac{q^3(q^3-p^3)+m^2}{\epsilon_{q^3}\epsilon_{q^3-p^3}}\right)\frac{\theta(\tilde\mu-\epsilon_{q^3-p^3})+\theta(\tilde\mu-\epsilon_{q^3})}{\epsilon_{q^3-p^3}+\epsilon_{q^3}}\right],
\end{align}
and now taking the limit $p^3\rightarrow 0$, we obtain in the leading order, 
\begin{align}
\Pi^{00}_{\rm med}(p^0=0, p^3\rightarrow 0)= 2N_ce_f^2 |e_fB|\int_0^\infty\frac{dq^3}{(2\pi)^2}\frac{\partial \theta(\tilde\mu-\epsilon_{q^3})}{\partial \epsilon_{q^3}}=-\frac{N_ce_f^2 |e_fB|}{2\pi^2}\frac{\tilde\mu}{\sqrt{\tilde\mu^2-m^2}}\theta(\tilde\mu-m).\label{Debyemass}
\end{align}

Note that, the same result for the Debye mass can be obtained alternatively from the second order derivative of the thermodynamic potential (\ref{thermo-pot}) with respect to the temporal component of the external electromagnetic field $A^0$. See Appendix B for details. 

It is interesting to check that the polarization tensor, in the strong field approximation, satisfies the transversality condition in the reduced $(1+1)$-D space, i.e., $p^{\mu\parallel}\Pi_{\mu\nu}^\parallel(p^\parallel)=\Pi_{\mu\nu}^\parallel(p^\parallel)p^{\nu\parallel}=0$. In statistical QED in the presence of a magnetic field, the photon polarization tensor has nine independent gauge invariant tensorial structures \cite{Hugo}. At a strong magnetic field, however, there is only one independent structure given by $g_{\mu\nu}^\parallel-p^\parallel_\mu p^\parallel_\nu/(p^\parallel)^2$ due to the fact that the transverse momentum is zero for the quarks in the LLL \cite{Debye-MCFL} . Thus
\begin{equation}
\Pi_{\mu\nu}^f(p^\parallel)=\Pi(p^\parallel,\tilde\mu,e_fB)\left(g_{\mu\nu}^\parallel-\frac{p^\parallel_\mu p^\parallel_\nu}{(p^\parallel)^2}\right),\label{covariantform}
\end{equation}
with $\Pi(p^\parallel,\tilde\mu, e_fB)$ being a scalar coefficient, which can be obtained by contracting $\Pi^{\mu\nu}$ with the tensor $g_{\mu\nu}^\parallel$
\begin{align}
\Pi(p^\parallel,\tilde\mu, e_fB)=g^{\mu\nu\parallel} \Pi^f_{\mu\nu}=\Pi^f_{00}-\Pi^f_{33}. \label{coefficientquation}
\end{align}

From (\ref{00component}) and (\ref{33component}), we have
\begin{equation}
\Pi(p^\parallel,\tilde\mu, e_fB)=4N_ce_f^2 \frac{|e_fB|}{\beta}\sum_{q^4}\int\frac{dq^3}{(2\pi)^2}\frac{1}{\left[(iq^4-\tilde\mu-ip^4)^2-\epsilon_{q^3-p^3}^2\right][(iq^4-\tilde\mu)^2-\epsilon_{q^3}^2]}.
\end{equation}
Carrying out the sum of the Matsubara frequencies and making the continuation to Minkowski space, one has
\begin{align}
\nonumber\Pi(p^\parallel,\tilde\mu, e_fB)=&-N_ce_f^2 |e_fB|\int\frac{dq^3}{(2\pi)^2}\frac{1}{\epsilon_{q^3-p^3}\epsilon_{q^3}}\left[\frac{n_F(\epsilon_{q^3-p^3}+\tilde\mu)+n_F(\epsilon_{q^3}-\tilde\mu)}{p^0+\epsilon_{q^3-p^3}+\epsilon_{q^3}} + \frac{n_F(\epsilon_{q^3-p^3}-\tilde\mu)-n_F(\epsilon_{q^3}-\tilde\mu)}{p^0-\epsilon_{q^3-p^3}+\epsilon_{q^3}}\right.\\
&+\left.\frac{n_F(\epsilon_{q^3}+\tilde\mu)-n_F(\epsilon_{q^3-p^3}+\tilde\mu)}{p^0+\epsilon_{q^3-p^3}-\epsilon_{q^3}} - \frac{n_F(\epsilon_{q^3}+\tilde\mu)+n_F(\epsilon_{q^3-p^3}-\tilde\mu)}{p^0-\epsilon_{q^3-p^3}-\epsilon_{q^3}}\right]+\Pi_{\rm vaccum}.
\end{align}
At zero temperature and  in the infrared limit ($p^0=0, p^3\rightarrow 0$), only the statistical part contributes,
\begin{align}
\nonumber \Pi(p^\parallel=0,\tilde\mu, e_fB)=&-2N_ce_f^2 |e_fB|\int_{m}^\infty\frac{d\epsilon_{q^3}}{(2\pi)^2}\frac{1}{\epsilon_{q^3}\sqrt{\epsilon_{q^3}^2-m^2}}\left[\frac{\theta(\tilde\mu-\epsilon_{q^3})}{\epsilon_{q^3}} - \frac{\partial\theta(\tilde\mu-\epsilon_{q^3})}{\partial\epsilon_{q^3}}\right],\\
=&-\frac{N_ce_f^2 |e_fB|}{2\pi^2}\frac{\tilde\mu}{\sqrt{\tilde\mu^2-m^2}}\theta(\tilde\mu-m),
\end{align}

From (\ref{covariantform}), we thus have
\begin{align}
	\Pi^f_{00}(p^0=0, p^3\rightarrow 0)=\Pi(p^\parallel=0,{\tilde{\mu}},e_fB)=-\frac{N_ce_f^2 |e_fB|}{2\pi^2}\frac{\tilde\mu}{\sqrt{\tilde\mu^2-m^2}}\theta(\tilde\mu-m),\label{coefficient}
\end{align}	
which is exactly (\ref{Debyemass}). The result (\ref{coefficient}) also indicates that the $33-$component of the polarization operator,  $\Pi^f_{33}$, which appears in (\ref{coefficientquation}),
vanishes in the infrared limit.  This conclusion is demonstrated by explicit calculations in Appendix A. More physical implications of the vanishing of $\Pi^f_{33}$ in the infrared limit will be presented in subsection D.

\subsection{Debye Length}

The shielding of an external electric field in a plasma can be seen as a consequence of a dielectric polarization of the medium. That is, under an applied electric field, it takes place a redistribution of space charge that prevents the penetration of the external field beyond certain length. The length-scale associated with such shielding is the Debye length. 

The medium polarization is quantized by the 00-component of the photon polarization tensor in the infrared limit. Its effect is reflected in the modification of the Poisson equation, which for a point charge $q$ is then given by
 \begin{equation}\label{Poisson-Eq}
{\bm\nabla}^2A_0=-\Pi_{00} A_0-\frac{q}{\epsilon_0}\delta(r),
\end{equation} 
with $\epsilon_0$ the vacuum permittivity. The solution of (\ref{Poisson-Eq}) is the modified Coulomb law, 
\begin{equation}\label{Coulomb-Law}
A_0=\frac{q}{\epsilon_0r}e^{-r/\lambda_D},
\end{equation} 
with the Debye length given by $\lambda_D=1/\sqrt{-\Pi_{00}}$.

Hence, the screened Coulomb potential (\ref{Coulomb-Law}) can be seen as the one with the bare $1/r$ decay in the presence of a modified permittivity, $\epsilon$, which absorbs the medium polarization
\begin{equation}\label{Coulomb-Law-2}
A_0=\frac{q}{\epsilon r}, \quad  \epsilon=\epsilon_0e^{r/\lambda_D}.
\end{equation} 

That is, when $\Pi_{00}$ is not zero in the infrared limit, the Coulomb potential of a point charge $q$ is shielded on distance scales longer than $\lambda_D$ by a shielding cloud of approximate radius $\lambda_D$ consisting of charges with opposite sign. 

In particular, in the strongly magnetized MDCDW phase, we find from (\ref{Debyemass}) that the corresponding Debye length is
\begin{align}\label{Debye-Length}
\nonumber\lambda_D=&\frac{1}{\sqrt{-\sum_f\Pi_f^{00}}}=\left(\frac{2\pi^2}{N_c\sum_fe_f^2|e_fB|}\sqrt{1-\frac{m^2}{\tilde\mu^2}}\right)^{1/2}\\
=&\left(\frac{\pi}{2\alpha |eB|}\sqrt{1-\frac{m^2}{\tilde\mu^2}}\right)^{1/2},     \quad  \tilde\mu=| \mu -b| \geqslant m,
\end{align}
with $\alpha$ the fine-structure constant and $e$ the electron charge. In (\ref{Debye-Length}), we used the fact that $N_c=3$.
\begin{figure}
	\includegraphics[height=13cm]{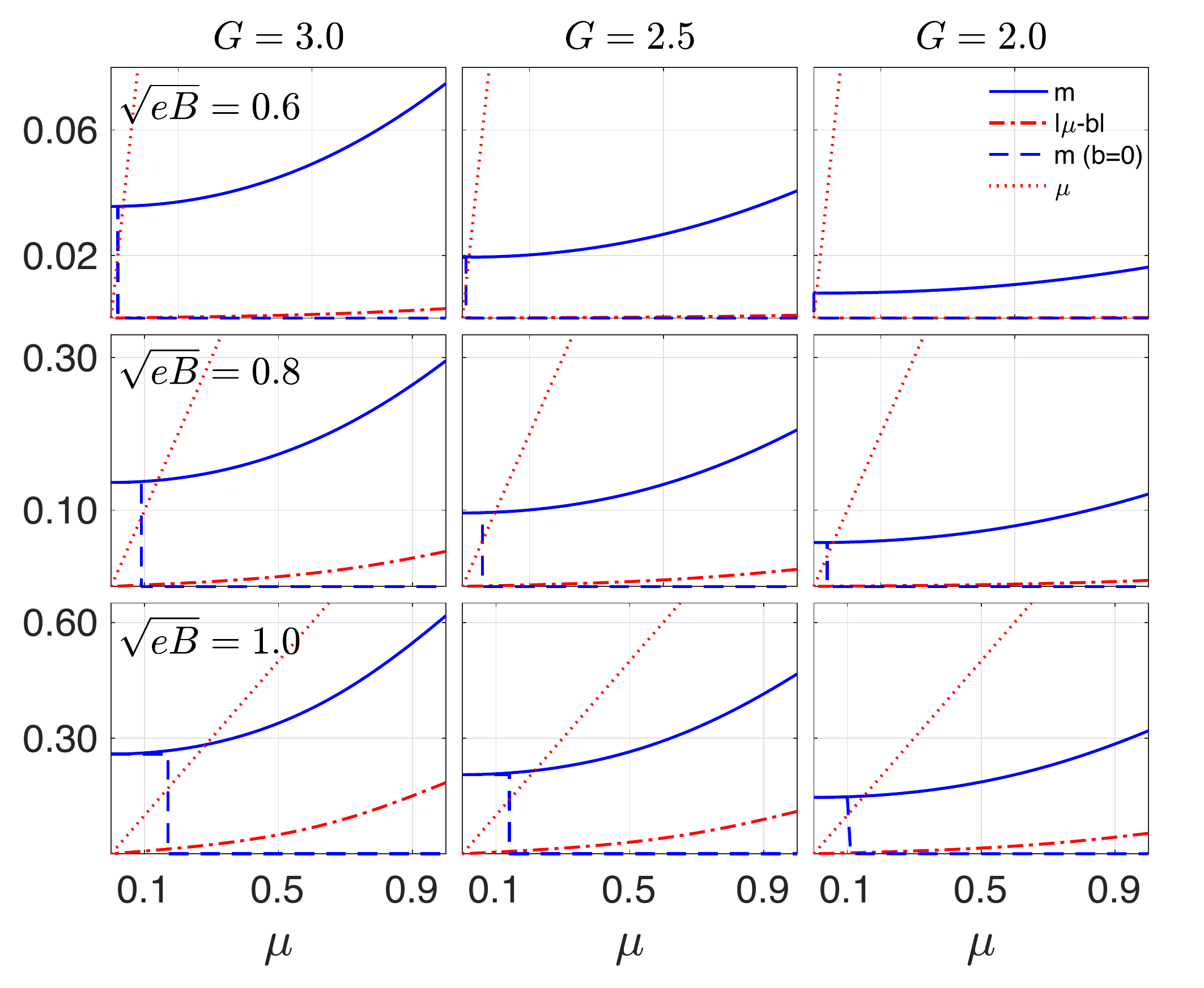}
	\caption{(Color online) Parameters' values in the density region at magnetic-field values in the horizontal rows  ($\sqrt{eB}=0.6, 0.8, 1.0$) and coupling constants ($G= 2.0, 2.5, 3.0$) in the vertical columns. The solutions are obtained in the LLL approximation. }\label{Parameter-Domain}
\end{figure}

In the MDCDW phase, the Debye length depends on the dynamical parameters $m$ and $b$, so they have to be found for each value of chemical potential through the gap equations (\ref{Gap-Eq}). Now, from (\ref{Debye-Length}), we have that $\lambda_D$ will be different from zero only if $| \mu -b| > m$. But, from Fig. \ref{Parameter-Domain} we see that $| \mu -b| < m$ in the density domain where the LLL-approximation is valid. Thus, in this phase, there will be no Debye screening in that density region in the strong-magnetic field range $0.6 \leqslant \sqrt{eB} \leqslant 1.0$. 

In the case of MC$\chi$SB at finite density, the Debye screening requires that $\mu > m$. Thus, we find that it will take place only beyond the critical chemical potential value for the phase transition to the chiral symmetric phase with $m=0$. However, since as shown in Section 3, the more favored phase at those densities is the MDCDW phase, we have that no Debye screening will take place in cold quark matter in a strong magnetic field at least in the density region where our approximation is valid. To extend our result to the whole domain of chemical potentials, it is needed to find the quark propagator under an arbitrary magnetic field, which will depend on all LL's, from where to calculate the corresponding photon polarization operator. This is an involved task that goes beyond the scope of the present work. Nevertheless, if we accept a $20\%$ of accuracy in our results, for $\sqrt{eB}\sim 1.0$ the lack of Debye screening will take place for a relatively wide range of densities ($0.2 < \mu < 1$).

\subsection{Meissner mass}

In our case, the Meissner mass is obtained from the $33$-component of the polarization operator (\ref{photonselfenergy}) in the infrared limit. As in the $00$-component case, the result after the Matsubara sum will be decomposed into the vacuum and statistical parts 
\begin{equation}
\Pi^{33}_f=\Pi^{33}_{\rm vac}+\Pi^{33}_{\rm med}.
\end{equation}
The details of the calculation will be given in the Appendix A. Here, we give the results after the infrared limit is carried out
\begin{align}
	\Pi^{33}_{\rm vac}(p^0=0, p^3 \rightarrow 0)
	=N_ce_f^2 |e_fB|\sum_{s=0}^SC_s\int\frac{dq^3}{(2\pi)^2}\frac{M_s^2}{\epsilon^3_{q^3}}=N_ce_f^2 |e_fB|\frac{1}{2\pi^2}\sum_{s=0}^SC_s=0,
\end{align}
where we take into account the regularization constraint $\sum_{s=0}^SC_s=0$.

For the statistical part at zero temperature 
\begin{align}
\Pi^{33}_{\rm med}(p^0=0, p^3 \rightarrow 0)=&N_ce_f^2 |e_fB|\int\frac{dq^3}{(2\pi)^2}\frac{(q^3)^2}{(q^3)^2+m^2}\left[\frac{\partial \theta(\tilde\mu-\epsilon_{q^3})}{\partial\epsilon_{q^3}}+\frac{m^2}{(q^3)^2}\frac{1}{\epsilon_{q^3}}\theta(\tilde\mu-\epsilon_{q^3})\right]=0.\label{33medium}
\end{align}
Here, the two terms in the bracket cancel out. 

Therefore, there is no Meissner effect in this medium. This is a physically expected result, since the particle-hole condensate under consideration is electrically neutral and consequently the electromagnetic $U(1)$ symmetry is not broken by the ground state.

\section{Concluding remarks}

In this paper we study the screening effect of strongly magnetized cold quark matter in the intermediate density region. With this goal, we compared the phase characterized by a quark-antiquark chiral condensate (i.e. the MC$\chi$SB phase) with the one with a quark-hole inhomogeneous condensate (i.e. the MDCDW phase). We showed that at weak coupling the MDCDW phase is the more stable one at any density value. This result, together with the one reported in Ref. \cite{Klimenko} for strong coupling, indicates that the MDCDW phase is energetically favored for the entire coupling domain in the intermediate density region. This is an important result, since it changes the previous belief that a massless fermion theory under a strong magnetic field at finite density will dynamically generate a mass through a homogeneous chiral condensate (i.e. the generation of the MC$\chi$SB phase) that will be present up to certain critical value of the chemical potential \cite{Leung}, from where on the condensate is evaporated and the chiral symmetry is reinstated. By the present result, we have that this homogeneous phase will be unstable at any $\mu$ value, and that it should be replaced by a more energetically favored phase characterized by an inhomogeneous particle-hole condensate (i.e. the MDCDW phase). 

Another significant part of this paper has been devoted to the investigation of the screening effects on this dense medium.  In order to find the Debye and Meissner masses, we calculated the photon polarization operator of the MDCDW phase at finite density and in the presence of a uniform and constant magnetic field in the one-loop approximation and in the strong-field limit. The $00$-component of the calculated polarization operator was different from zero under the condition that $| \mu -b| \geqslant m$ in the infrared limit ($p_0=0, {\bf p}\rightarrow 0$). Since $m$ and $b$ are dynamical parameters that should be found as the solution of the gap equations (\ref{Gap-Eq}), the Debye length becomes a dynamical quantity that will be different from zero only if the condition $| \mu -b| > m$ is satisfied. It happens that in the density region valid in the LLL-approximation, the condition $m > | \mu -b|$ leads, as it can be seen from the plots in Fig. \ref{Parameter-Domain}. Hence, in this quark matter phase no generated static electric field can be screened. Physically this can be understood from the fact that in this phase all the particles are forming neutral pairs, thus, no charge exists to form a screening cloud.

On the other hand, there is no Meissner effect in this medium (i.e. $\Pi^{33}(p^0=0, p^3 \rightarrow 0)=0$). This is physically expected by the same reason previously mentioned, that is, because the particle-hole condensate under consideration is electrically neutral and consequently the electromagnetic $U(1)$ symmetry is not broken by the ground state.

This result puts a finishing touch on the screening effects of cold dense quark matter under a strong magnetic field: if it is considered that in moving from the lower to the higher density region quark matter passes from the MDCDW to the CFL color-superconducting phase respectively, we have that the electric screening, as well as the magnetostatic screening (i.e. Meissner effect) will be absent in the relatively high-density region.  We recall that the MCFL  phase of color superconductivity lacks of Debye and Meissner screening \cite{Debye-MCFL}, since all the quarks in this phase are paired in neutral Cooper pairs with respect to the in-medium electromagnetism \cite{Cristina}.   

Our findings can be of interest for the astrophysics of NS, where the conditions of relatively high densities and strong magnetic fields prevail. On the other hand, as we already pointed out in the Introduction, the study of electric field effects on astronomical objects has been a topic of investigation since many years ago  \cite{E-NS}.
The  topic  has got  a  renewed attention due to the fact that many observed pulsars have been estimated to have huge electric fields in their surfaces \cite{Sahakian}. Moreover, it is noteworthy that stars' collapse to a singularity point  may be avoided by the existence of a net charge in its composition that can counter balance the gravitational attraction by the Coulomb repulsion if it is not screened. On the same footing, a charged compact star can be in hydrodynamical equilibrium having a radius on the verge of forming an event horizon.
Thus, charged stars can exhibit larger masses for a given radius. Several mechanisms to charge the star has been already proposed. For example, in binary neutron-star systems, one of the components may acquire a net charge by accretion from the companion \cite{Accreation}. It has also been pointed out \cite{Usov} that in ultra-compact stars like ‘strange stars’ composed of u, d and s quarks, the electric field at the surface could be as high as $10^{19}$ eV/cm. In this context, to know that the different plausible phases of magnetized quark matter, if realized in the core of NS, do not have the capability to screen a generated electric field is a noteworthy result, since the presence of a Debye screening in the stellar medium will erase all these effects. The same is valid regarding the lack of magnetic screening in the whole density domain. As known, a strong star inner magnetic field can affect significantly its equation of state for nuclear matter \cite{NM-B}, as well as for quark-matter \cite{Paulucci, QM-B}.

\begin{acknowledgments}
EJF work was supported in part by NSF grant PHY-1714183. BF's work was supported in part by the NSF of China under grant no. 11535005.

\end{acknowledgments}

\appendix

\section{Other components of the photon polarization operator}

In Section IV, we calculated in detail only the $00-$component of the one-loop photon polarization operator in the strong field limit. In this Appendix, we shall present the calculation of the remaining components, especially the $33-$component that indicates the possible magnetostatic screening of this medium.  From (\ref{photonselfenergy}), we have

\begin{align}
\Pi_f^{33}(p^\parallel)=2N_ce_f^2 \frac{|e_fB|}{\beta}\sum_{q_4}\int\frac{dq^3}{(2\pi)^2}\frac{(iq^4-\tilde\mu)(iq^4-\tilde\mu-ip^4)+q^3(q^3-p^3)-m^2}{\left[(iq^4-\tilde\mu-ip^4)^2-\epsilon_{q^3-p^3}^2\right][(iq^4-\tilde\mu)^2-\epsilon_{q^3}^2]},\label{33component}
\end{align}
and
\begin{align}
	\Pi_f^{03}(p^\parallel)=	\Pi_f^{30}(p^\parallel)=2N_ce_f^2 \frac{|e_fB|}{\beta}\sum_{q_4}\int\frac{dq^3}{(2\pi)^2}\frac{(iq^4-\tilde\mu)(q^3-p^3)+(iq^4-\tilde\mu-ip^4)q^3}{\left[(iq^4-\tilde\mu-ip^4)^2-\epsilon_{q^3-p^3}^2\right][(iq^4-\tilde\mu)^2-\epsilon_{q^3}^2]}.\label{03component}
\end{align}
Carrying out the sum in Matsubara frequencies and  making the analytic continuation from Euclidean to Minkowski space, one obtains for the $03-$component
\begin{equation}
\Pi_f^{03}= \Pi^{03}_{\rm vac}+ \Pi^{03}_{\rm med},
\end{equation}
with
\begin{align}
	\Pi^{03}_{\rm vac}(p^\parallel)
	=N_ce_f^2 |e_fB|\sum_{s=0}^SC_s\int\frac{dq^3}{(2\pi)^2}\frac{p^0\left [(q^3-p^3)\epsilon_{q^3,M_s}-q^3\epsilon_{q^3-p^3,M_s}\right ]}
	{\left (\epsilon_{q^3,M_s}\epsilon_{q^3-p^3,M_s}\right) \left [(p^0)^2-(\epsilon_{q^3,M_s}+\epsilon_{q^3-p^3,M_s})^2\right ]},	
\end{align}
and
\begin{align}
	\nonumber \Pi^{03}_{\rm med}(p^\parallel)=& -2N_ce_f^2|e_fB|\int\frac{dq^3}{(2\pi)^2}\\
	\nonumber &\times\left\{\frac{q^3\epsilon_{q^3-p^3}+(q^3-p^3)\epsilon_{q^3}}{4\epsilon_{q^3}\epsilon_{q^3-p^3}}\left[\frac{n_F(\epsilon_{q^3}+\tilde\mu)-n_F(\epsilon_{q^3-p^3}+\tilde\mu)}{p^0-\epsilon_{q^3}+\epsilon_{q^3-p^3}}+\frac{n_F(\epsilon_{q^3}-\tilde\mu)-n_F(\epsilon_{q^3-p^3}-\tilde\mu)}{p^0+\epsilon_{q^3}-\epsilon_{q^3-p^3}} \right]\right.\\
	&+\left.\frac{q^3\epsilon_{q^3-p^3}-(q^3-p^3)\epsilon_{q^3}}{4\epsilon_{q^3}\epsilon_{q^3-p^3}}\left[\frac{n_F(\epsilon_{q^3}-\tilde\mu)+n_F(\epsilon_{q^3-p^3}+\tilde\mu)}{p^0+\epsilon_{q^3}+\epsilon_{q^3-p^3}}+\frac{n_F(\epsilon_{q^3}+\tilde\mu)+n_F(\epsilon_{q^3-p^3}-\tilde\mu)}{p^0-\epsilon_{q^3}-\epsilon_{q^3-p^3}} \right]\right\}.
\end{align}

Similarly, we have for the $33-$component
\begin{equation}
\Pi_f^{33}= \Pi^{33}_{\rm vac}+ \Pi^{33}_{\rm med},
\end{equation}
with
\begin{align}
\Pi^{33}_{\rm vac}(p^\parallel)
=-N_ce_f^2 |e_fB|\sum_{s=0}^SC_s\int\frac{dq^3}{(2\pi)^2}\left(\frac{1}{\epsilon_{q^3,M_s}}+\frac{1}{\epsilon_{q^3-p^3,M_s}}\right)\frac{\epsilon_{q^3,M_s}\epsilon_{q^3-p^3,M_s}-q^3(q^3-p^3)+M_s^2}{(p^0)^2-(\epsilon_{q^3,M_s}+\epsilon_{q^3-p^3,M_s})^2},
\end{align}
and
\begin{align}
\nonumber \Pi^{33}_{\rm med}(p^\parallel)=& 2N_ce_f^2|e_fB|\int\frac{dq^3}{(2\pi)^2}\\
\nonumber &\times\left\{\frac{1}{4}\left(1+\frac{q^3(q^3-p^3)-m^2}{\epsilon_{q^3}\epsilon_{q^3-p^3}}\right)\left[\frac{n_F(\epsilon_{q^3-p^3}+\tilde\mu)-n_F(\epsilon_{q^3}+\tilde\mu)}{p^0-\epsilon_{q^3}+\epsilon_{q^3-p^3}}-\frac{n_F(\epsilon_{q^3-p^3}-\tilde\mu)-n_F(\epsilon_{q^3}-\tilde\mu)}{p^0+\epsilon_{q^3}-\epsilon_{q^3-p^3}} \right]\right.\\
&+\left.\frac{1}{4}\left(1-\frac{q^3(q^3-p^3)-m^2}{\epsilon_{q^3}\epsilon_{q^3-p^3}}\right)\left[\frac{n_F(\epsilon_{q^3-p^3}+\tilde\mu)+n_F(\epsilon_{q^3}-\tilde\mu)}{p^0+\epsilon_{q^3}+\epsilon_{q^3-p^3}}-\frac{n_F(\epsilon_{q^3-p^3}-\tilde\mu)+n_F(\epsilon_{q^3}+\tilde\mu)}{p^0-\epsilon_{q^3}-\epsilon_{q^3-p^3}} \right]\right\}.
\end{align}
Notice that, we had regularized the vacuum parts by PV regularization. In the following, we will calculate the contribution from the vacuum and statistical parts in the static limit $p^0=0,p^3\rightarrow 0$. We thus again write the statistical part in a form in which the static limit is easy to handle. For the vacuum part,  we obtain
\begin{align}
	\Pi^{03}_{\rm vac}(p^0=0,p^3)=0,
\end{align}
and
\begin{align}
\Pi^{33}_{\rm vac}(p^0=0,p^3)
=N_ce_f^2 |e_fB|\sum_{s=0}^SC_s\int\frac{dq^3}{(2\pi)^2}\left[\frac{M_s^2}{\epsilon^3_{q^3}}+\frac{p^3q^3(M_s^2-2(q^3)^2)}{2\epsilon_{q^3,M_s}^5}+{\cal O}((p^3)^2)\right].
\end{align}
In the leading order, 
\begin{align}
	\Pi^{33}_{\rm vac}(p^\parallel=0)
	=N_ce_f^2 |e_fB|\sum_{s=0}^SC_s\int\frac{dq^3}{(2\pi)^2}\frac{M_s^2}{\epsilon^3_{q^3}}=N_ce_f^2 |e_fB|\frac{1}{2\pi^2}\sum_{s=0}^SC_s=0.
\end{align}
For the statistical part, at zero temperature, we have
\begin{align}
	\nonumber \Pi^{03}_{\rm med}(p^0=0,p^3)= -\frac{N_ce_f^2|e_fB|}{2}\int\frac{dq^3}{(2\pi)^2}
	&\left[ \left( \frac{q^3\epsilon_{q^3-p^3}+(q^3-p^3)\epsilon_{q^3}}{\epsilon_{q^3}\epsilon_{q^3-p^3}}\right )\frac{\theta(\tilde\mu-\epsilon_{q^3})-\theta(\tilde\mu-\epsilon_{q^3-p^3})}{\epsilon_{q^3}-\epsilon_{q^3-p^3}} \right.\\
	+&\left.\left (\frac{q^3\epsilon_{q^3-p^3}-(q^3-p^3)\epsilon_{q^3}}{\epsilon_{q^3}\epsilon_{q^3-p^3}}\right )\frac{\theta(\tilde\mu-\epsilon_{q^3})-\theta(\tilde\mu-\epsilon_{q^3-p^3})}{\epsilon_{q^3}+\epsilon_{q^3-p^3}}\right],
\end{align}
and
\begin{align}
\nonumber \Pi^{33}_{\rm med}(p^0=0, p^3)=&\frac{N_ce_f^2 |e_fB|}{2}\int\frac{dq^3}{(2\pi)^2}\left[\left(1+\frac{q^3(q^3-p^3)-m^2}{\epsilon_{q^3}\epsilon_{q^3-p^3}}\right)\frac{\theta(\tilde\mu-\epsilon_{q^3-p^3})-\theta(\tilde\mu-\epsilon_{q^3})}{\epsilon_{q^3-p^3}-\epsilon_{q^3}} \right.\\
&\ \ \ \ \ \ \ \ \ \ \ \ \ \ \ \ \ \ \ \ \ \ \  +\left.\left(1-\frac{q^3(q^3-p^3)-m^2}{\epsilon_{q^3}\epsilon_{q^3-p^3}}\right)\frac{\theta(\tilde\mu-\epsilon_{q^3-p^3})+\theta(\tilde\mu-\epsilon_{q^3})}{\epsilon_{q^3-p^3}+\epsilon_{q^3}}\right].
\end{align}
Taking the limit $p^3\rightarrow 0$, they become
\begin{align}
\Pi^{03}_{\rm med}(p^0=0, p^3 \rightarrow 0)= -N_ce_f^2|e_fB|\int\frac{dq^3}{(2\pi)^2}
	\frac{q^3}{\epsilon_{q^3}}\frac{\partial\theta(\tilde\mu-\epsilon_{q^3})}{\partial\epsilon_{q^3}}=0,
\end{align}
and

\begin{align}
\nonumber \Pi^{33}_{\rm med}(p^0=0, p^3 \rightarrow 0)=&N_ce_f^2 |e_fB|\int\frac{dq^3}{(2\pi)^2}\frac{(q^3)^2}{(q^3)^2+m^2}\left[\frac{\partial \theta(\tilde\mu-\epsilon_{q^3})}{\partial\epsilon_{q^3}}+\frac{m^2}{(q^3)^2}\frac{1}{\epsilon_{q^3}}\theta(\tilde\mu-\epsilon_{q^3})\right]\\
\nonumber =&-\frac{N_ce_f^2 |e_fB|}{2\pi^2}\left[\int_m^\infty\frac{d\epsilon_{q^3}}{\epsilon_{q^3}}\sqrt{\epsilon_{q^3}^2-m^2}\delta(\tilde\mu-\epsilon_{q^3})-\int_m^{\tilde\mu}\frac{d\epsilon_{q^3}}{\epsilon_{q^3}^2}\frac{m^2}{\sqrt{\epsilon_{q^3}^2-m^2}}\right]\\
\nonumber =& -\frac{N_ce_f^2 |e_fB|}{2\pi^2}\left[\frac{\sqrt{\tilde\mu^2-m^2}}{\tilde\mu}-\frac{\sqrt{\tilde\mu^2-m^2}}{\tilde\mu}\right] \\
=&0.\label{33medium}
\end{align}

Therefore, as expected, there is no Meissner effect in the medium (i.e. $\Pi^{33}(p^0=0, p^3 \rightarrow 0)=0$) since the particle-hole condensate under consideration is electrically neutral and consequently the electromagnetic $U(1)$ symmetry is not broken by the ground state. This results is also consistent with the covariant form of the photon polarization operator that we claimed in (\ref{covariantform}).

\section{An alternative way to obtain the Debye mass}

As known, the Debye mass can also be obtained from the second-order derivative of the thermodynamic potential (\ref{thermo-pot}) with respect to the temporal component of the external electromagnetic field $A^0$, which is related to the chemical potential via the simple relation, $e_fA^0\rightarrow \mu$. At zero temperature, the only part contributing to the Debye mass is the medium one, i.e., $\Omega_\mu$. In LLL, it reads \cite{E-V-PLB}
\begin{align}
	\nonumber\Omega^{fLLL}_\mu=&-\frac{1}{2}\frac{N_c|e_fB|}{(2\pi)^2}\int_{-\infty}^\infty dp^3\sum_{\epsilon}\left(|E_0-\mu|-E_0\right)\\
\nonumber 	=&-\frac{N_c|e_fB|}{(2\pi)^2}\left\{\left[Q(\mu)+m^2\ln\frac{m}{R(\mu)} \right]\theta(b-\mu-m)\theta(b-m)\right.\\
	\nonumber &-\left[Q(0)+m^2\ln\frac{m}{R(0)}\right]\theta(b-m)+\left[Q(\mu)+m^2\ln\frac{m}{R(\mu)}\right]\theta(\mu-b-m)\\
	&-\left. \left[Q(0)+m^2\ln\frac{m}{R(0)}\right]\theta(\mu-b-m)\theta(-b-m)\right\},
\end{align}
with
\begin{align}
	Q(\mu)=|b-\mu|\sqrt{(b-\mu)^2-m^2}, \ \ \ \ \ \ \ R(\mu)=|b-\mu|+\sqrt{(b-\mu)^2-m^2}.
\end{align}
One thus obtains
\begin{align}
	\frac{\partial^2 \Omega^{fLLL}_\mu}{\partial\mu^2}=-\frac{N_c|e_fB|}{2\pi^2}\frac{\tilde\mu}{\sqrt{\tilde\mu^2-m^2}}\theta(\tilde\mu-m).
\end{align}

Therefore, we obtain the Debye mass by replacing $A^0$ with $\mu/e_f$ 
\begin{align}
\frac{\partial^2 \Omega^{fLLL}_\mu}{\partial(A^0)^2}=e_f^2\frac{\partial^2 \Omega^{fLLL}_\mu}{\partial\mu^2}=-\frac{N_ce_f^2|e_fB|}{2\pi^2}\frac{\tilde\mu}{\sqrt{\tilde\mu^2-m^2}}\theta(\tilde\mu-m).
\end{align}
This is exactly the result we obtained in (\ref{Debyemass}).

\newpage 

\end{document}